\documentstyle[twoside,fleqn,espcrc2,epsf]{article}

\title{Analysis of Hadron Propagators with One Thousand Configurations
on a $24^3\times 64$ Lattice at $\beta=6.0$\thanks{presented by A. Ukawa}}

\author{
JLQCD Collaboration \\[2mm]
S. Aoki\address{Institute of Physics, University of Tsukuba, Tsukuba,
Ibaraki 305, Japan},
M. Fukugita\address{Yukawa Institute for Theoretical Physics,
Kyoto University, Kyoto 606,
Japan},
S. Hashimoto\address{National Laboratory for High Energy Physics (KEK),
Tsukuba, Ibaraki 305, Japan},
Y. Iwasaki$^{\rm a,}$\address{Center for Computational Physics,
University of Tsukuba, Tsukuba, Ibaraki 305, Japan},
K. Kanaya$^{\rm a,d}$,
Y. Kuramashi$^{\rm c}$,
H. Mino\address{Faculty of Engineering, Yamanashi University, Kofu 540,
Japan},\\
M. Okawa$^{\rm c}$,
A. Ukawa$^{\rm a}$,
T. Yoshi\'e$^{\rm a,d}$}

\begin{document}

\begin{abstract}
Statistical properties of effective mass are analyzed.  We show from a
general ground that effective mass as a function of time should not exhibit
long plateaux whatever high statistics simulations are made: the mass should
fluctuate beyond the one standard deviation of error bars after a few time
slices for large times where the ground state dominates.
This explains the difficulty of obtaining long plateaux experienced in
previous simulations.
Implications of the observation for global $\chi^2$ fits
are discussed, and results for hadron masses are presented.
\end{abstract}

\maketitle

\section{Introduction}

Calculation of hadron masses constitutes a basic part in virtually all
problems of
lattice QCD simulations.  A quantity always examined in such calculations
is the
effective mass $m_{\rm eff}(t)$. The existence of a plateau in
$m_{\rm eff}(t)$ as a
function of time $t$ is regarded as a dual measure for
the statistical quality of
data and the minimum time separation beyond which the ground state
dominates. In this way the plateau is used as a guide in the choice of
the time interval for a
global $\chi^2$ fitting to extract hadron masses from propagators.
In practice,
however, a long plateau has rarely been seen.
Even in the best previous efforts
toward high statistics simulations\cite{ape,qcdpax,gf11,lanl},
effective masses,
particularly for $\rho$ meson and the nucleon, almost always deviate
from a plateau beyond error bars
after 5 or 6 times slices.
Ensuing uncertainties in the choice of the fitting range and fitted values
of hadron masses have represented a
severe hindrance factor in attempts toward high precision determination of
hadron
masses, especially for light hadrons\cite{ukawa}.

In this report we present an analysis on the origin of this problem.
Our study is based on hadron propagators
for the Wilson quark action at $K=0.1545, 0.1550, 0.1555$ evaluated on
1000 quenched
gauge configurations on a $24^3\times 64$ lattice at $\beta=6.0$, which were
 generated with the
5-hit pseudoheatbath algorithm at 2000 sweep intervals.  The central value
$K=0.1550$ has been chosen to facilitate a comparison with previous
high statistics
studies\cite{ape,qcdpax,lanl} which employ up to 400 configurations
for the same spatial
size\cite{qcdpax} or a larger size of $32^3$\cite{lanl}.

\begin{figure*}[t]
\centerline{\epsfxsize=160mm \epsfbox{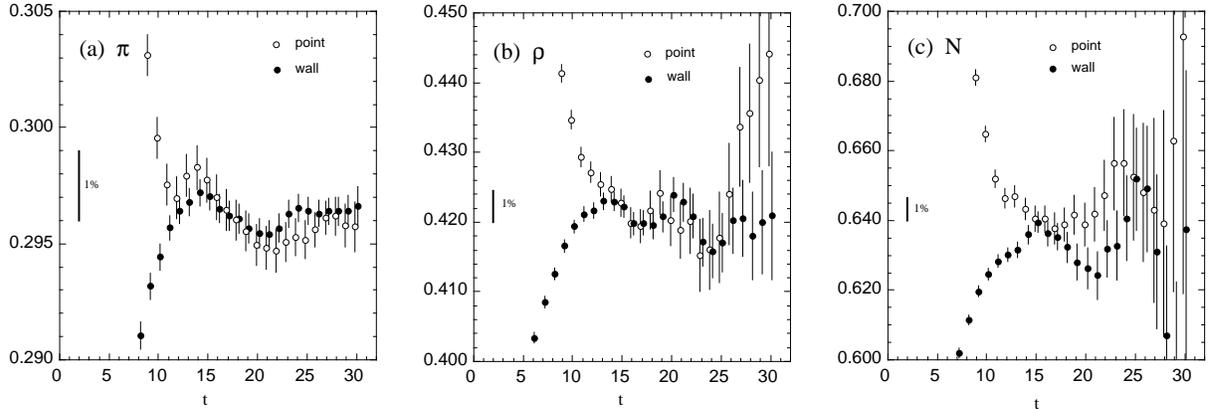}}
\vspace*{-1cm}
\caption{Effective mass for (a) $\pi$, (b) $\rho$ and (c) $N$ at $\beta=6.0$
and $K=0.1550$ on a $24^3\times 64$ lattice with 1000 configurations. Errors
are estimated by a single elimination jackknife procedure.}
\label{fig:fig1}
\vspace*{-3mm}
\end{figure*}

This calculation is one of the
first QCD runs carried out by the JLQCD Collaboration on VPP500/80 at KEK which
started operation in January 1995.
The machine consists of 80 processing elements (PE's),
each with the peak speed of 1.6GFLOPS and 256MBytes of memory,  connected by a
crossbar switch.  Our run used 64 processors, on which our code for heatbath
and red/black minimal residual solver
sustained the speed of $1.0-1.3$ GFLOPS/PE.  The configuration generation of
2000 sweeps and a calculation of standard meson and baryon propagators for the
point and wall sources on the final configuration was made in about 30
minutes, so that the entire run took about 20 days to complete.

\section{Analysis of effective mass}

In Fig.~\ref{fig:fig1} we show the effective mass for $\pi$,
$\rho$ and nucleon ($N$) obtained with the point(open circles) or the
wall (filled
circles) source at $K=0.1550$ or $m_\pi/m_\rho\approx 0.7$,
which roughly corresponds to strange quark mass.
A striking feature observed in these plots is the presence of fluctuations
which exceed the level of one standard deviation after a few time slices
even for $\pi$ and with 1000 configurations.

In fact it is possible to understand this behavior by a simple statistical
analysis.  To show this, let
$\overline{G(t)}$ be the average for a hadron propagator over $N$ independent
configurations, and $G_{\rm true}(t)=\langle G(t)\rangle$ be the
true propagator.  According to the central
limit theorem, the difference $\delta G(t)=\overline{G(t)}-G_{\rm true}(t)$
obeys
the
distribution, \begin{equation} P[\overline{G}]\propto \exp\left[
-\frac{N}{2}\sum_{t,t'} \delta G(t)C^{-1}(t,t')\delta G(t')\right],
\label{eq:distribution} \end{equation}
where the covariance matrix $C$ is defined by
\begin{equation}
C(t,t')=\langle G(t)G(t')\rangle-\langle G(t)\rangle\langle G(t')\rangle.
\end{equation}
Let us introduce a normalized covariance matrix
\begin{equation}
\tilde C(t,t')\equiv\frac{C(t,t')}{\sqrt{C(t,t)C(t',t')}}
\end{equation}
and denote the eigenvalues and normalized
eigenvectors of $\tilde C(t,t')$ by $\lambda_i$ and $e_i(t)$ $(i=1, \cdots,
L_t)$ with $L_t$ the temporal lattice size.
An eigenvector decomposition of the
averaged propagator leads to the formula,
\begin{equation}
\overline{G(t)}=G_{\rm true}(t)+\sqrt{\frac{C(t,t)}{N}}\sum_{i=1}^{L_t}
\sqrt{\lambda_i}e_i(t)\xi_i,
\label{eq:eigendecom}
\end{equation}
where, according to (\ref{eq:distribution}), $\xi_i\, (i=1, \cdots, L_t)$ are
independent Gaussian random numbers.

\begin{figure*}
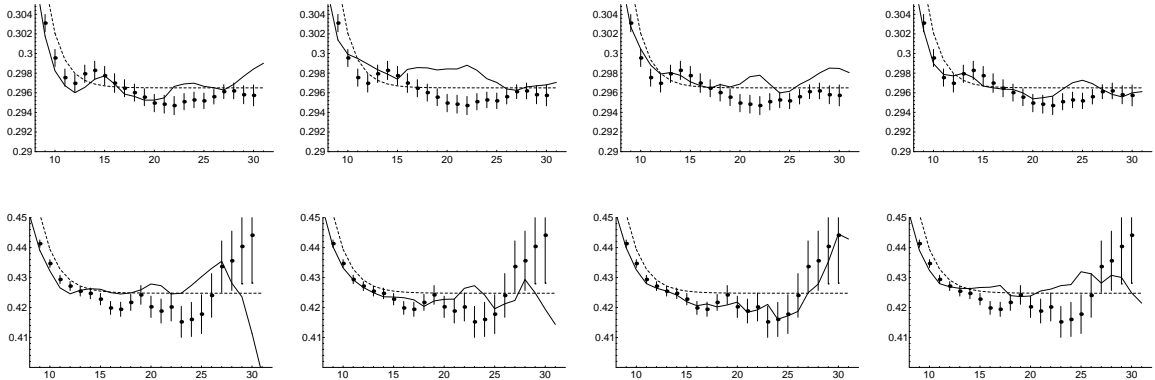

%\vspace*{5cm}
%\centerline{
\epsfxsize=160mm \epsfbox{fig2a.epsf}
%}
\vspace*{5mm}
\centerline{\epsfxsize=160mm \epsfbox{fig2b.epsf}}
\vspace*{-0.5cm}
\caption{Four examples of effective mass for $\pi$ (top row) and
$\rho$ (bottom row)
for point source calculated from ``simulated'' propagators (solid curves)
as compared to the measured values for $K=0.1550$ (shown with error bars)
taken from Fig.~\protect\ref{fig:fig1}.
Dotted lines correspond to the model (\protect\ref{eq:model}).}
\label{fig:fig2}
\end{figure*}

Given a model of the true propagator $G_{\rm true}(t)$ and a measured value
of the covariance matrix, this formula allows us to generate ``simulated''
samples of
the averaged propagator $\overline{G(t)}$ by generating a set of $L_t$
Gaussian random
numbers $\xi_i$.

For the model of $G_{\rm true}(t)$ we take a double hyperbolic
cosine form,
\begin{equation}
G_{\rm true}(t)=Ze^{-mt}+Z'e^{-m't}+(t\to L_t-t),
\label{eq:model}
\end{equation}
where the masses and residues are determined by a $\chi^2$ fit of
the measured propagator over $6\leq t\leq L_t/2$. In Fig.~\ref{fig:fig2} we
show four examples of effective mass for $\pi$ and $\rho$ for the point source
calculated from ``simulated'' propagators (solid curves).
Dotted lines represent values
for the model (\ref{eq:model}).  These examples clearly demonstrate that
fluctuations as observed
in Fig.~\ref{fig:fig1} are a typical occurrence.
Indeed some of the examples are
amusingly similar to the measured value.

Let us add a remark that the diagonal of the
covariance matrix is expected to behave as $C(t,t)\propto\exp (-\alpha t)$ with
$\alpha=2m_\pi$ for $\pi$ and $\rho$ and  $\alpha=3m_\pi$ for $N$\cite{lepage},
with which our data are consistent.  Combined with (\ref{eq:eigendecom})
this explains why the magnitude of fluctuation of $m_{\rm eff}$ increases
rapidly for $\rho$
and $N$ toward large times, while it stays roughly constant for $\pi$.

We now reexamine the concept of a plateau in the light of the above analysis.
Let us define a plateau over the time interval $t_1\leq t\leq t_2$ by
the condition
that $m_{\rm eff}(t)$ over this interval falls within a band of width
$2\delta m$
centered at $m$, {\it i.e.,}  $m-\delta m\leq
m_{\rm eff}(t)\leq m+\delta m$ for $t=t_1, \cdots, t_2$. To calculate
the probability
for the occurrence of such a plateau,
we make a change of variable $P[\overline{G}]\to P[m_{\rm eff}(t_1),\cdots,
m_{\rm eff}(t_2)]$ in (\ref{eq:distribution}), defining
$m_{\rm eff}(t)=\log\overline{G(t)}/\overline{G(t+1)}$ (effects of the
periodic boundary condition are small in the numerical results below).
The probability is then given by an
integral of
$P[m_{t_1},\cdots, m_{t_2}]$ over $m-\delta m\leq m_{t_i}\leq m+\delta
m (i=t_1,\cdots, t_2)$.

\begin{figure}[t]
\centerline{\epsfxsize=74mm \epsfbox{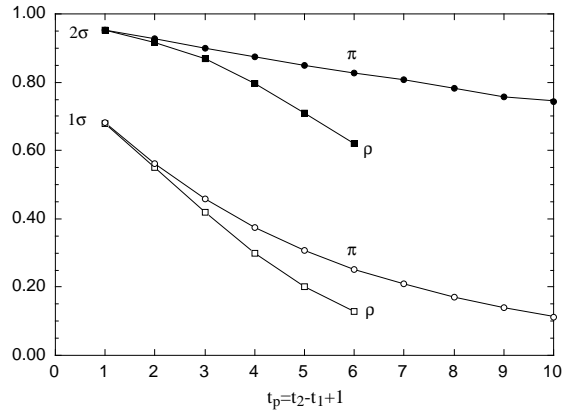}}
\vspace{-1cm}
\caption{Probability for plateau of length $t_p$ starting at $t_1=15$ for
$\pi$ and
$\rho$.  Open and filled symbols correspond to the choice of $\delta m$
equal to one or two
standard deviations of $m_{\rm eff}(t_1=15)$.}
\label{fig:fig3}
\vspace*{-10mm}
\end{figure}

In Fig.~\ref{fig:fig3} we plot the probability of finding a plateau of length
$t_p(=t_2-t_1+1)$ for $\pi$ and $\rho$ calculated with the measured
covariance matrix for $K=0.1550$ for the case of the wall source.
We fix $t_1=15$, since we have chosen the ground state mass $m$ by a
$\chi^2$ fit with a single hyperbolic cosine  over $15\leq t\leq L_t/2$.  For
$\delta m$ we choose one (open symbols) or two (filled symbols) standard
deviation of $m_{\rm eff}(t)$ at $t=t_1$.  Values for
$\rho$ for $t\geq 7$ are not shown because of the poor quality of the
covariance matrix for large times.

This figure shows that one
{\it should not} expect to see a plateau at the level of one standard
deviation even for
$\pi$: the probability drops below 50\% after 2 time slices.  Allowing for
a deviation of two standard
deviations, a plateau of length 10 for $\pi$ ($3-4$ units in terms of
the correlation
length $\xi_\pi=1/m_\pi$) becomes probable, while for $\rho$ the probability
decreases to $60-50$\% already at $t\approx 6-7$ ($t/\xi_\rho\approx 3$).
These features are consistent with actual examples of effective
mass shown in Fig.~\ref{fig:fig1}, and also with those of previous high
statistics
simulations\cite{ape,qcdpax,gf11,lanl}.

We should emphasize that the pattern of fluctuations of effective mass does
not change
when one increases the number of configurations $N$, except that the magnitude
scales down as $1/\sqrt{N}$.  At the same time statistical errors
estimated for $m_{\rm eff}$ also decreases as $1/\sqrt{N}$.  In this sense
higher
statistics does not lead to a longer or better plateau.  In other words we can
ask for such an improvement only within a fixed magnitude of the
absolute error ({\it e.g.,} 1\% of mass).  Let us add that the use of larger
spatial volumes and improved operators having a larger coupling to
the ground state
also help to obtain a better plateau only in the latter sense.

\section{$\chi^2$ fits for hadron masses}

\begin{table}
\setlength{\tabcolsep}{0.2pc}
\caption{Hadron masses in lattice units at $\beta=6.0$.  For our results and
those of QCDPAX the first and second row correspond to point and wall source.
APE used multi-origin $7^3$ source, while LANL results are combined estimates
from wall and Wuppertal-smeared source.}
\label{tab:tab1}
\vspace*{3mm}
\begin{tabular}{llll}
\hline
$\quad K$&$\quad\pi$&$\quad\rho$&$\quad N$\\
\hline
0.1545&0.33075(40)&0.4441(6) &0.6828(21)\\
      &0.33076(28)&0.4425(10)&0.6777(21)\\
0.1550&0.29642(42)&0.4231(16)&0.6451(11)\\
      &0.29642(27)&0.4220(12)&0.6393(27)\\
0.1555&0.25867(46)&0.4019(20)&0.6066(13)\\
      &0.25864(33)&0.4016(17)&0.6003(37)\\
\hline
\multicolumn{4}{l}{Previous results at $K=0.1550$}\\
\hline
APE&0.298(2)&0.429(3)&0.647(6)\\
QCDPAX&0.2960(8)&0.4201(29)&0.6403(50)\\
      &0.2964(6)&0.4228(19)&0.6307(39)\\
LANL&0.297(1)&0.422(3)&0.641(4)\\
\hline
\end{tabular}
\vspace*{-5mm}
\end{table}

Our analysis should have made it clear that restricting the fitting range of
a global $\chi^2$ fit to the time interval of an apparent plateau is not well
founded.  In fact if one repeats a simulation with a different sequence
of random
numbers the effective mass will generally exhibit a plateau at some
other time interval at a different value.
This indicates that it is more reasonable to take the minimum time $t_{min}$
at the time slice where the dominance of the ground state is reasonably ensured
({\it e.g.,} by the overlap of $m_{\rm eff}$ for the point and wall sources),
and to
extend the fitting range to large times as long as statistical fluctuations
do not
become unacceptably large, without resorting to the presence of plateau.
The value of correlated $\chi^2$ should tell whether
the choice is reasonable.  Of course the fitted values of hadron
masses vary depending on the choice of the interval.  However, this is an
uncertainty which can only be reduced by an improved measurement of
hadron propagators.

In Table~\ref{tab:tab1} we present hadron masses obtained by a correlated
$\chi^2$ fit over the interval $15\leq t\leq L_t/2=32$
with a single hyperbolic cosine
for $\pi$ and $\rho$ and with a single exponential for $N$.
Errors correspond to an increase of $\chi^2$
by one.  For $\pi$ and $\rho$ our results obtained with the point and
wall sources
are mutually in agreement.  For $K=0.1550$ previous results from the QCDPAX
Collaboration\cite{qcdpax} and from the Los Alamos group\cite{lanl}
are consistent
with ours, while those from APE are higher.  The case of nucleon is
problematical.  We find that the $\chi^2$ fit is not very
stable and that the point and wall results do not agree within the error.
Effort
toward improving baryon operators will be needed for a precise determination of
baryon masses even in the region of strange quark.

\end{document}